\documentstyle[aps,12pt]{revtex}
\begin{document}
\title{Solution of relativistic Hartree-Bogoliubov equations in\\
configurational representation: spherical neutron halo nuclei}
\author{M. Stoitsov$^{1,2}$, P. Ring$^{2}$, D. Vretenar$^{2,3}$ and G.A. 
Lalazissis$%
^{2}$}
\address{$^{1}$Institute for Nuclear Research and Nuclear Energy, \\
Bulgarian Academy of Sciences, 
Sofia--1784, Bulgaria\\
$^{2}$ Physik-Department der Technischen Universit\"at M\"unchen,
D-85748 Garching, Germany\\
$^{3}$ Physics Department, Faculty of Science, University of
Zagreb, Croatia\\
\symbol{126}}
\maketitle
\begin{abstract}
The scaled harmonic oscillator basis (SHO) is derived by
a local scaling-point transformation of the spherical
harmonic oscillator radial wave functions. The unitary scaling
transformation produces a basis with  
improved asymptotic properties. The SHO basis is employed
in the solution of the relativistic Hartree-Bogoliubov (RHB) 
equations in configurational space. The model is applied in the 
self-consistent mean-field approximation to the description 
of the neutron halo in Ne isotopes. It is shown that an 
expansion of nucleon spinors and mean-field potentials
in the SHO basis reproduces the asymptotic properties of 
neutron densities calculated by finite element discretization
in the coordinate space. In the RHB description of 
neutron skins and halos, SHO bases in two or three
dimensions can be a useful alternative to technically complicated
solutions on a mesh in coordinate space.
\end{abstract}

\section{Introduction}

Theoretical models based on the relativistic mean-field approximation have 
been successfully applied in the description of a variety of nuclear 
structure phenomena in spherical and deformed $\beta$-stable nuclei~\cite
{Rin.96}, and more recently in studies of exotic nuclei far from the 
valley of beta stability~\cite{PVL.97,LVP.98,LVR.97,VPLR.98,LVR.98,VLR.98}.
Short lived isotopes with extreme neutron - to - proton ratios 
display unique properties. On the neutron-rich side in particular, 
extremely diffuse and spatially extended neutron densities are found in 
nuclei close to the particle drip lines. Densities with large spatial 
extensions result from the extremely weak binding of the outermost 
nucleons and the coupling between bound states and the particle 
continuum. The theoretical description of the appearance of neutron
skins, or eventually neutron halos, necessitates a careful treatment 
of the asymptotic part of the nucleonic densities. While calculations 
based on modern Monte Carlo shell-model techniques provide an accurate
description of the structure of relatively light exotic nuclei, for 
medium-heavy and heavy systems the only viable approach at present are 
mean-field models. 

In addition to the self-consistent mean-field potentials, 
pairing correlations have to be taken into account in order 
to describe ground-state properties of open shell nuclei. 
For strongly bound 
systems pairing can be included in the form of a simple BCS approximation 
in the valence shell. For drip line nuclei on the other hand, the separation
energy of the last nucleons can become extremely small. The Fermi level
is found close to the particle continuum, and the 
lowest particle-hole or particle-particle modes are embedded in the
continuum. The BCS pairing model presents only a poor approximation
and predicts unbound systems.  A correct description 
of the scattering of nucleonic pairs from bound states to the
positive energy continuum, and therefore of the formation of
spatially extended nucleonic densities, necessitates a unified 
treatment of mean-field and pairing correlations. In the non 
relativistic framework the structure of spherical exotic neutron-rich, 
as well as proton drip-line nuclei, has been described in the 
Hartree-Fock-Bogoliubov (HFB) theory in coordinate
space~\cite{DNW.96,Naz.96}. Among models based on
quantum hadrodynamics, the relativistic Hartree-Bogoliubov (RHB) 
model~\cite{LVP.98,KR.91} has been very successfully employed
in calculations of ground-state properties of spherical exotic nuclei 
on both sides of the valley of $\beta$-stability. 
For $\beta$-stable nuclei the traditional
Hartree-Fock models and the relativistic mean-field theory
predict very similar ground-state densities. On the other hand,
recent applications to the structure 
of drip-line nuclei have shown significant differences,
especially in the isospin dependence of the spin-orbit
potential. 

In Ref.~\cite{LVP.98} it was emphasized that for a 
correct theoretical description of nuclei 
with large neutron excess, the model must include: 
(i) a self-consistent solution for the
mean-field in order to obtain the correct increase of
the surface diffuseness, (ii) the correct isospin dependence of the 
spin-orbit term in the mean-field potential, (iii) a unified 
self-consistent description of pairing correlations, (iv) coordinate
space solutions for the equations that describe the
coupling between bound and continuum states. However, the coordinate 
space representation of HFB or RHB
equations still presents serious difficulties
in the calculation of ground-state properties of deformed nuclei.
For finite-range interactions 
the technical and numerical problems on a two-dimensional mesh 
are so involved that reliable self-consistent 
calculations in coordinate space should not be expected soon, and 
therefore one has to look for alternative temporary solutions.
In the configurational representation the HFB or RHB systems of partial 
differential equations have been solved by expanding the nucleon wave
functions and mean-field potentials in a large oscillator basis.
In many applications an expansion of the wave functions in an
appropriate harmonic oscillator (HO) basis of spherical or axial symmetry 
provides a satisfactory level of accuracy.
For nuclei at the drip lines the expansion in the localized
oscillator basis presents only a poor approximation for
the continuum states. Oscillator expansions produce densities which
decrease to steeply in the asymptotic region at large distance from the
center of the nucleus.
The calculated rms radii cannot reproduce the 
experimental values, especially for halo nuclei.

In a recent work~\cite{SNP.98} a new configurational representation basis 
has been suggested, based on a unitary transformation of the
spherical HO basis. The new basis is derived from
the standard oscillator basis by performing a local-scaling point 
coordinate transformation~\cite{Pet83a}. The precise form of the 
transformation
is determined by the desired asymptotic behavior of the densities.
At the same time it preserves many useful properties of the HO wave 
functions.
Using the new basis, characteristics of weakly bound orbitals for 
a square-well potential were analyzed, and the ground-state 
properties of some spherical nuclei were calculated in the framework
of the energy density functional approach. It was shown that 
properties of spatially extended states can be described 
using a rather modest number of basis states. The new basis, it was 
suggested, might present an interesting alternative to 
the algorithms that are being developed for the coordinate space solutions 
of
the HFB equations. In the present work we investigate a particular 
scaled HO basis which might be appropriate for the solution
of the coupled system
of Dirac-Hartree-Bogoliubov integro-differential eigenvalue equations,
and Klein-Gordon equations for the meson fields. 

In Sec.~2 we present an outline of the relativistic
Hartree-Bogoliubov theory and discuss the various
approximations. The local--scaling transformed spherical 
HO basis is introduced in Sec.~3, and results of calculations 
in the configurational 
representation for a series of Ne isotopes are compared with 
solutions obtained by discretization of the system of RHB
equations on the finite element mesh in coordinate space
\cite{PVL.97,PVR2.97}. A summary of our results is presented in Sec.~4.


\section{Relativistic Hartree-Bogoliubov Model}

The relativistic Hartree-Bogoliubov theory provides a unified
description of mean-field and pairing correlations in nuclei. 
In this section we briefly outline the essential characteristics 
of the model that will be used in the calculation of ground-state
properties. A detailed description of the model 
can be found in Refs.~\cite{LVP.98,PVR2.97}.
The ground state of a nucleus $\vert \Phi >$ is described
as vacuum with respect to independent quasi-particle operators.
The generalized single-nucleon Hamiltonian 
contains two average potentials: the self-consistent mean-field
$\hat\Gamma$ which encloses all the long range particle-hole ({\it ph})
correlations, and a pairing field $\hat\Delta$ which sums
up the particle-particle ({\it pp}) correlations. The variation of the 
energy functional
with respect to the hermitian density matrix $\rho$
and the antisymmetric pairing tensor $\kappa$, 
produces the single quasi-particle eigenvalue equations. 
The model was formulated
in Ref.~\cite{KR.91}. In the Hartree approximation for
the self-consistent mean field, the relativistic
Hartree-Bogoliubov (RHB) equations read
\begin{eqnarray}
\label{equ.2.2}
\left( \matrix{ \hat h_D -m- \lambda & \hat\Delta \cr
                -\hat\Delta^* & -\hat h_D + m +\lambda} \right) 
\left( \matrix{ U_k({\bf r}) \cr V_k({\bf r}) } \right) =
E_k\left( \matrix{ U_k({\bf r}) \cr V_k({\bf r}) } \right).
\end{eqnarray}
where $\hat h_D$ is the single-nucleon Dirac
Hamiltonian and $m$ is the nucleon mass
\begin{equation}
\hat{h}_{D}=-i\mathbf{\alpha \cdot \nabla }+\beta (m+g_{\sigma }\sigma
(r))+g_{\omega }\tau _{3}\omega ^{0}(r)+g_{\rho }\rho ^{0}(r)+e{\frac{{%
(1-\tau _{3})}}{2}}A^{0}(r).  \label{dirh}
\end{equation}
The chemical potential $\lambda$  has to be determined by
the particle number subsidiary condition in order that the
expectation value of the particle number operator
in the ground state equals the number of nucleons. The column
vectors denote the quasi-particle spinors and $E_k$
are the quasi-particle energies. The Dirac Hamiltonian 
contains the mean-field potentials of the isoscalar 
scalar $\sigma$-meson, the isoscalar vector $\omega$-meson,
the isovector vector $\rho$-meson, as well as the 
electrostatic potential. 
The RHB equations have to be solved self-consistently with
potentials determined in the mean-field approximation from
solutions of Klein-Gordon equations.
The equation for the sigma meson contains the 
non-linear $\sigma$ self-interaction terms~\cite{BB.77}.
Because of charge conservation only the
three-component of the isovector $\rho$-meson contributes. The
source terms for the Klein-Gordon equations are calculated 
in the {\it no-sea} approximation.

The pairing field
$\hat\Delta $ in (\ref{equ.2.2}) reads
\begin{equation}
\label{equ.2.5}
\Delta_{ab} ({\bf r}, {\bf r}') = {1\over 2}\sum\limits_{c,d}
V_{abcd}({\bf r},{\bf r}') {\bf\kappa}_{cd}({\bf r},{\bf r}'),
\end{equation}
where $a,b,c,d$ denote quantum numbers
that specify the single-nucleon states, 
$V_{abcd}({\bf r},{\bf r}')$ are matrix elements of a
general two-body pairing interaction, and the pairing
tensor is defined 
\begin{equation}
{\bf\kappa}_{cd}({\bf r},{\bf r}') = 
\sum_{E_k>0} U_{ck}^*({\bf r})V_{dk}({\bf r}').
\end{equation}
The matrix elements in the pairing channel $V_{abcd}$  can in principle 
be derived as a one-meson exchange interaction by eliminating
the mesonic degrees of freedom in the model Lagrangian~\cite{KR.91}.
However, the resulting  pairing matrix elements are unrealistically large 
if for the coupling constants the standard parameter sets 
of the mean-field approximation are used.
These parameters do not reproduce scattering data
On the other hand, since we are using effective forces, 
there is no fundamental reason to have the same interaction both 
in the particle-hole and particle-particle channel.  
Although encouraging results have been reported in applications 
to nuclear matter, a microscopic and fully relativistic derivation 
of the pairing force starting from the Lagrangian of quantum
hadrodynamics still cannot be applied to realistic nuclei.
Therefore,  it was suggested in Ref.~\cite{Gonz.96} that an appropriate
approximation might be a phenomenological Gogny-type 
finite range interaction
\begin{eqnarray}
V^{pp}(1,2)~=~\sum_{i=1,2}
e^{-( ({\bf r}_1- {\bf r}_2)
 / {\mu_i} )^2}\,
(W_i~+~B_i P^\sigma \nonumber \\
-H_i P^\tau -
M_i P^\sigma P^\tau),
\end{eqnarray}
with the parameters
$\mu_i$, $W_i$, $B_i$, $H_i$ and $M_i$ $(i=1,2)$. The procedure requires no 
cut-off and provides a very reliable description of pairing properties in 
finite nuclei.

The eigensolutions of Eq. (\ref{equ.2.2}) form a set of
orthogonal and normalized single quasi-particle states. The corresponding
eigenvalues are the single quasi-particle energies.
The self-consistent iteration procedure is performed
in the basis of quasi-particle states. The resulting quasi-particle
eigenspectrum is then transformed into the canonical basis of single-particle
states, in which the RHB ground-state takes the  
BCS form. The transformation determines the energies
and occupation probabilities of the canonical states.

The self-consistent solution of the 
Dirac-Hartree-Bogoliubov integro-differential eigenvalue equations
and Klein-Gordon equations for the meson fields determines the 
nuclear ground state. In Refs.~\cite{PVL.97,LVP.98,LVR.97,VPLR.98}
we used finite element methods
in the coordinate space discretization of the coupled system
of equations. The model was applied to exotic neutron-rich nuclei.
Very diffuse ground-state densities were described and the 
occurrence of neutron halo was investigated. In the description
of not so exotic Ni and Sn isotopes~\cite{LVR.98}, as well as
for proton drip line nuclei~\cite{VLR.98}, we followed the  
prescription of Ref.~\cite{Gonz.96}:
the Dirac-Hartree-Bogoliubov equations and the equations for the 
meson fields were solved by expanding the nucleon spinors 
$U_k({\bf r})$ and $V_k({\bf r})$, 
and the meson fields in a basis of spherical harmonic oscillators 
for $N = 20$ oscillator shells. For nuclei at the proton drip 
lines we verified the results by comparing calculations in 
configuration space with coordinate space solutions. Differences
were negligible. As will be shown in the next section, things
are very different on the neutron-rich side.


\section{Calculations in configurational representation}



\subsection{The scaled harmonic oscillator basis}


In the construction of the local--scaling transformed harmonic 
oscillator basis we will use
the local-scaling transformation method \cite{Pet83a}.
A local-scaling point coordinate
transformation (LST) is defined as
\begin{equation}
{\bf r}^{\prime }={\bf f}({\bf r})\equiv \hat{\bf r}f({\bf r)}.
\end{equation}
The transformed radius vector has the same direction 
$\hat{\bf r} \equiv {\bf r}/|{\bf r}|$, while its 
magnitude $r^{\prime }=f({\bf r)}$ depends on the scalar LST function.
$f({\bf r)}$ is assumed to be an increasing function of $r$,
and $f(\bf{0})$=0. The set of invertible transformations of this type forms a
LST group. The corresponding LST wave function can be expressed as 
\begin{equation}\label{lstwf} 
\Psi_{f}({\bf r}_{1},{\bf r}_{2},...,{\bf r}_{A})=\left[ \prod_{i=1}^{A}
\frac{f^2({\bf r}_i)}{r^2_i}\frac{\partial f({\bf r}_i)}{\partial r_i}
\right]^{1/2}\bar{\Psi}({\bf f}({\bf r}_{1}),{\bf f}({\bf r}_{2}),...,{\bf f}
({\bf r}_{A})),  
\end{equation} 
where $\bar{\Psi}({\bf r}_{1},{\bf r}_{2},...,{\bf r}_{A})$ is a 
model $A$-particle wave function
normalized to unity 
\begin{equation} 
\langle \bar{\Psi}|\bar{\Psi} \rangle =1.
\end{equation}
The transformation (\ref{lstwf}) is unitary and therefore the LST
wave function $\Psi_{f}({\bf r}_{1},{\bf r}_{2},...,{\bf r}_{A})$ is also
normalized to unity, independent of the choice of $f({\bf r})$.

The local one-body density corresponding to a $A$-body wave function  
${\Psi}$ is 
\begin{equation}
 {\rho}({\bf r})=A\int |{\Psi}
({\bf r},{\bf r}_{2},...,{\bf r}_{A})|^{2}
d{\bf r}_{2},...,d{\bf r}_{A}.  
\label{modden}
\end{equation} 
From the expression (\ref{lstwf}) it follows that there
exists a simple relation between the local density $\rho_f(\bf r)$ associated
with the LST function ${\Psi}_f$, and the density $\bar{\rho}(\bf r)$
which corresponds to the prototypical model function  $\bar{\Psi}$: 
\begin{equation}
\label{lstden} 
\rho_f({\bf r})=\frac{f^{2}({\bf r})}{r^{2}}\frac{\partial f({\bf
r})} {\partial r}\bar{\rho}(f({\bf r)}). 
\end{equation} 
This relation
is particularly useful when the form of the density $\rho_f(\bf r)$ 
is known, for example from experimental data.
In that case
Eq.~(\ref{lstden}) becomes a first order nonlinear differential
equation for the LST function $f$. For a system with spherical symmetry, 
$\rho_f$, $\bar{\rho}$, and $f$ depend only on  $r$=$|{\bf r|}$, and
Eq.~(\ref{lstden}) can be reduced to a nonlinear algebraic equation
\begin{equation}\label{lsteq}
\int\limits_{0}^{r}\rho_f
(u)u^{2}du=\int\limits_{0}^{f(r)}\bar{\rho}(u)u^{2}du.
\end{equation} 
The solution is found subject to the boundary condition $f(0)=0$.

For shell-model or mean-field applications we have to consider the
case when the model many-body wave function is a Slater determinant
\begin{equation}\label{ssmod} 
\bar{\Psi}({\bf r}_{1},{\bf r}_{2},...,{\bf r}_{A})=
\frac{1}{\sqrt{A!}}\det |\bar{\phi}_{i}({\bf r}_{j})|. 
\end{equation}
The single-particle functions $\bar{\phi}_{i}({\bf r})$
form a complete set. The LST wave function is defined by 
the transformation (\ref{lstwf}), and of course it is written
as a product state
\begin{equation}\label{ssf} 
\Psi_{f}({\bf r}_{1},{\bf r}_{2},...,{\bf r}_{A})=\frac{1}
{\sqrt{A!}}\det
|\phi_{i}({\bf r}_{j})| ,
\end{equation} 
of the LST single-particle basis states 
\begin{equation} \label{lstspwf}
\phi _{i}({\bf r})=\left[\frac{f^{2}({\bf r})}{r^{2}}
\frac{\partial f({\bf r})} 
{\partial r}\right]^{1/2}\bar{\phi}_{i}({\bf f}({\bf r)}). 
\end{equation} 

In the present work we consider systems with spherical symmetry. The  
single-particle basis reads
\begin{equation}
\bar{\phi} _{i}({\bf r,}\sigma ,\tau {\bf )=}R_{nl}^{{\rm HO}}(r){\cal J}%
_{jlm}(\Omega ,\sigma)\chi _{1/2m_{t}(\tau )}.  \label{howf}
\end{equation}
The angular part describes the coupling of orbital angular momentum and spin
\begin{equation}
{\cal J}_{i}(\Omega ,\sigma)\equiv J_{jlm_{j}}(\Omega ,\sigma)=
\sum_{m_{l}m_{s}}(lm_{l}1/2m_{s}|jm_{j})Y_{lm_{l}}(\Omega )\chi
_{1/2m_{s}(\sigma )}.  \label{sawf}
\end{equation}
The harmonic oscillator radial functions 
\begin{equation}
R_{nl}^{{\rm HO}}(r)=\frac{2(n-1)!b^{-3/2}}{\Gamma (n+l+1/2)^{3}}\left(
r/b\right) ^{l+1}e^{-\frac{1}{2}\left( \frac{r}{b}\right)
^{2}}L_{n-1}^{l+1/2}(r/b)  \label{horadwf}
\end{equation}
are characterized by a single parameter: the oscillator length 
$b=\sqrt{\hbar /m\omega }$.

If applied in the description of very weakly bound nuclei, the  
main drawback of HO wave functions is that, because of their Gaussian
asymptotics, they cannot describe the density profiles far away from the
center of the nucleus. In principle one could try to resolve this 
problem by including a very large number of oscillator shells, 
but extremely large bases are impossible to handle in 
more microscopic applications.
Therefore, by means of eqs (\ref{lstspwf}) and (\ref{howf}) it might be 
useful to define a scaled harmonic
oscillator (SHO) basis 
\begin{equation}
\phi _{i}({\bf r,}\sigma ,\tau {\bf )=}R_{nl}^{{\rm SHO}}(r){\cal J}%
_{jlm}(\Omega ,\sigma ,\tau )\chi _{1/2m_{t}(\tau )}.  \label{thowf}
\end{equation}
The angular part of course does not change,
while the radial functions $R_{nl}^{{\rm SHO}}(r)$ are 
defined by the local-scaling transformation 
\begin{equation}
R_{nl}^{{\rm SHO}}(r)=\left[ \frac{f^{2}(r)}{r^{2}}\frac{df(r)}{dr}\right]
^{1/2}R_{nl}^{{\rm HO}}(f(r)).  \label{thoradwf}
\end{equation}
The choice of the LST function 
\begin{equation}
f(r)=\left\{ 
\begin{array}{lll}
r & for & r\leq a \\ 
&  &  \\ 
a\left( 8r\,/a-{8}a/r+a^{2}/{r}^{2}-12\ln (r/a)\right) ^{1/2} & for & r>a
\end{array}
\right.  \label{lst}
\end{equation}
is motivated by the desired asymptotic form of the nucleonic densities.
For this particular parametrization the derivatives of $f(r)$ up to fourth 
order are continuous at the matching radius $a$, and 
$f(r)$ is a monotonically increasing function of $r$. 
The transformation (\ref{thoradwf}) is unitary; the Jacobian 
$\frac{f^{2}(r)}{r^{2}}\frac{df(r)}{dr}$ is positive for all values of $r$.
This makes the SHO basis complete,
since it is unitary equivalent to the complete HO basis.
The ansatz (\ref{lst}) for the LST function guarantees that all SHO basis
states (\ref{thowf}) are localized in space. 
Asymptotically, the linear term in the expansion (\ref{lst})
dominates and $f(r\rightarrow \infty )\sim r^{1/2}$. 
Therefore, at large distances 
the radial functions decrease exponentially. The SHO
functions are continuous (up to fourth derivatives) and depend on two
positive parameters: the oscillator parameter $b$ and the matching radius $a$.

It should be emphasized that the simple parametrization  (\ref{lst})
is motivated by our desire to keep the new basis as simple and practical 
as possible. 
In Ref.~\cite{LKR.96} it has been shown that a three-parameter form of the
LST function performs very well in actual calculations of ground-state 
properties of spherical nuclei. For light nuclei 
one of these parameters turns out to be very small. In the present application 
we choose it identically zero. The resulting matching conditions are
satisfied analytically and the parametrization (\ref{lst}) is derived.
Of course other parametrizations of $f(r)$
are possible. There are many ways to optimize the LST function,
depending on the actual physical problem under consideration.


\subsection{Neutron-rich Ne isotopes}


In Ref.~\cite{PVL.97} we have applied the RHB 
model to the description of the neutron halo in the mass region above
the s-d shell. Pairing correlations and the coupling to particle continuum
states have been described by finite range two-body Gogny-type interaction.
Finite element methods have been used in the coordinate space discretization
of the coupled system of Dirac-Hartree-Bogoliubov and Klein-Gordon equations.
As we have already emphasized, solutions in coordinate space
appear to be essential for the correct description of the coupling
between bound and continuum states. 
Using the parameter set NL3~\cite{LKR.96} for
the mean-field Lagrangian, and D1S~\cite{BGG.84}
parameters for the Gogny interaction, we have found
evidence for the occurrence of multi-neutron halos in
heavier Ne isotopes. We have shown that the
properties of the 1f-2p orbitals near the Fermi level and
the neutron pairing interaction play a crucial role in
the formation of the halo. In the present work we
essentially repeat the calculations of Ref.~\cite{PVL.97},
but in configurational representation with the SHO basis
defined by the LS transformation (\ref{thowf}).

In Fig.~\ref{figA} we plot the proton and neutron rms radii for the 
Ne isotopes as functions of the mass number. Results of both 
calculations in the coordinate space and in the configurational 
representation (HO and SHO bases) are shown. Proton radii change very 
little and only display a slow increase. There is practically 
no difference between results of coordinate and configurational 
space calculations. Neutron radii of Ne isotopes follow the 
mean-field  N$^{1/3}$ curve up to N $\approx$ 22. 
For larger values of N a sharp increase of neutron radii
is observed (calculations in coordinate space).
The matter radii of course follow this change and the 
spatial extension of heavier isotopes is substantially enlarged.
The sudden increase in neutron
rms radii has been interpreted as evidence for the
formation of a multi-particle halo~\cite{PVL.97}. We notice 
that neutron radii calculated with the expansion in the HO basis 
cannot reproduce the coordinate space results. 
Results calculated with 12 and 20 oscillator shells are compared,
and in both cases the radii follow the average mean-field curve.
The neutron radii increase with the number of oscillator shells, 
but their values are still far from the 
results obtained in coordinate space. 
On the other hand, radii
which result from the expansion in the SHO basis 
almost exactly reproduce values of finite element 
discretization in coordinate space. All calculations
with the SHO basis have been performed with 20 oscillator shells.
The same SHO basis has been used both for protons and for neutrons.

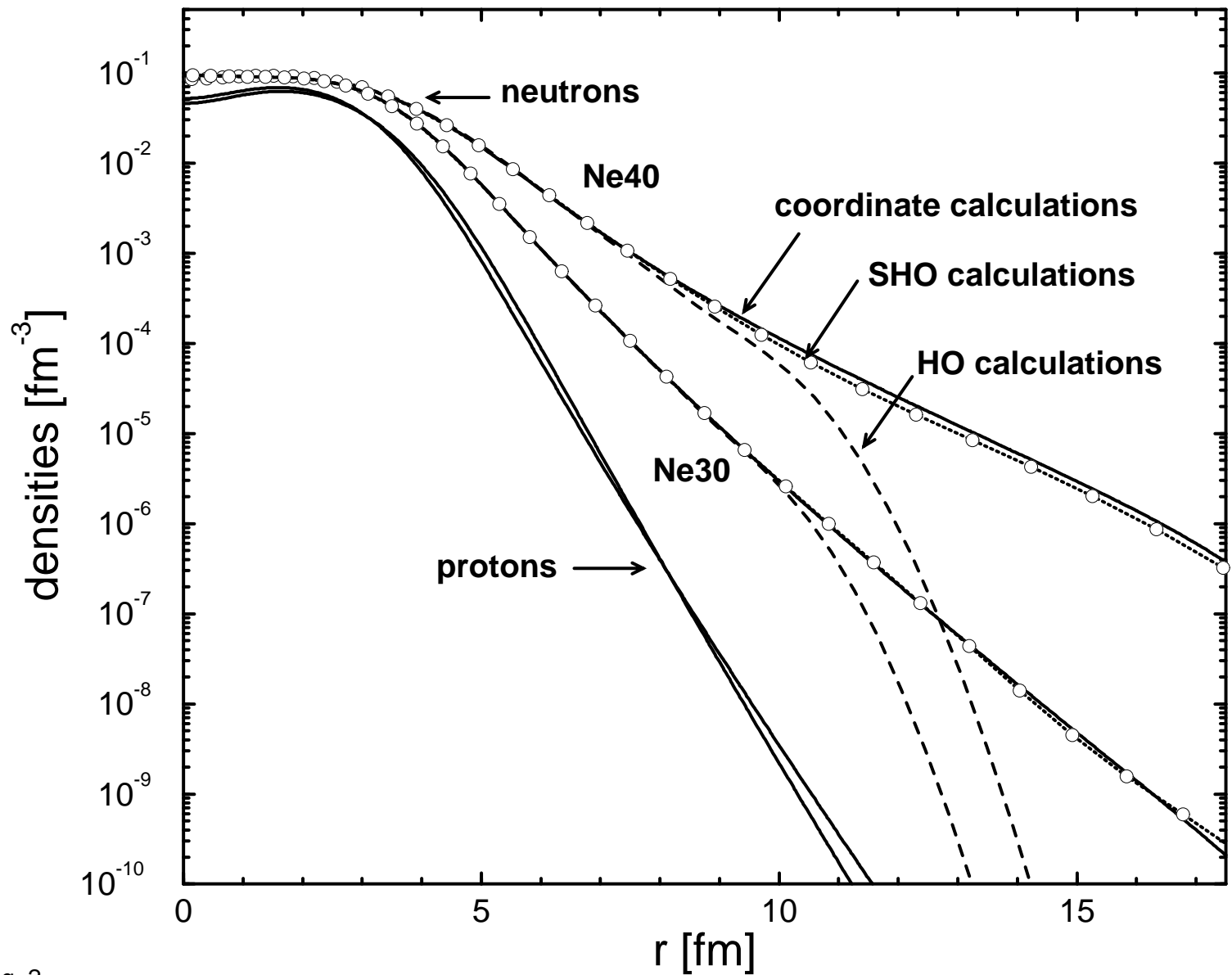
\begin{figure}
\caption{ Self-consistent RHB proton and neutron densities for the 
ground-states of $^{30}$Ne and $^{40}$Ne. Density profiles calculated 
in configurational space are compared with results on the 
coordinate mesh.}
\label{figB}
\end{figure}

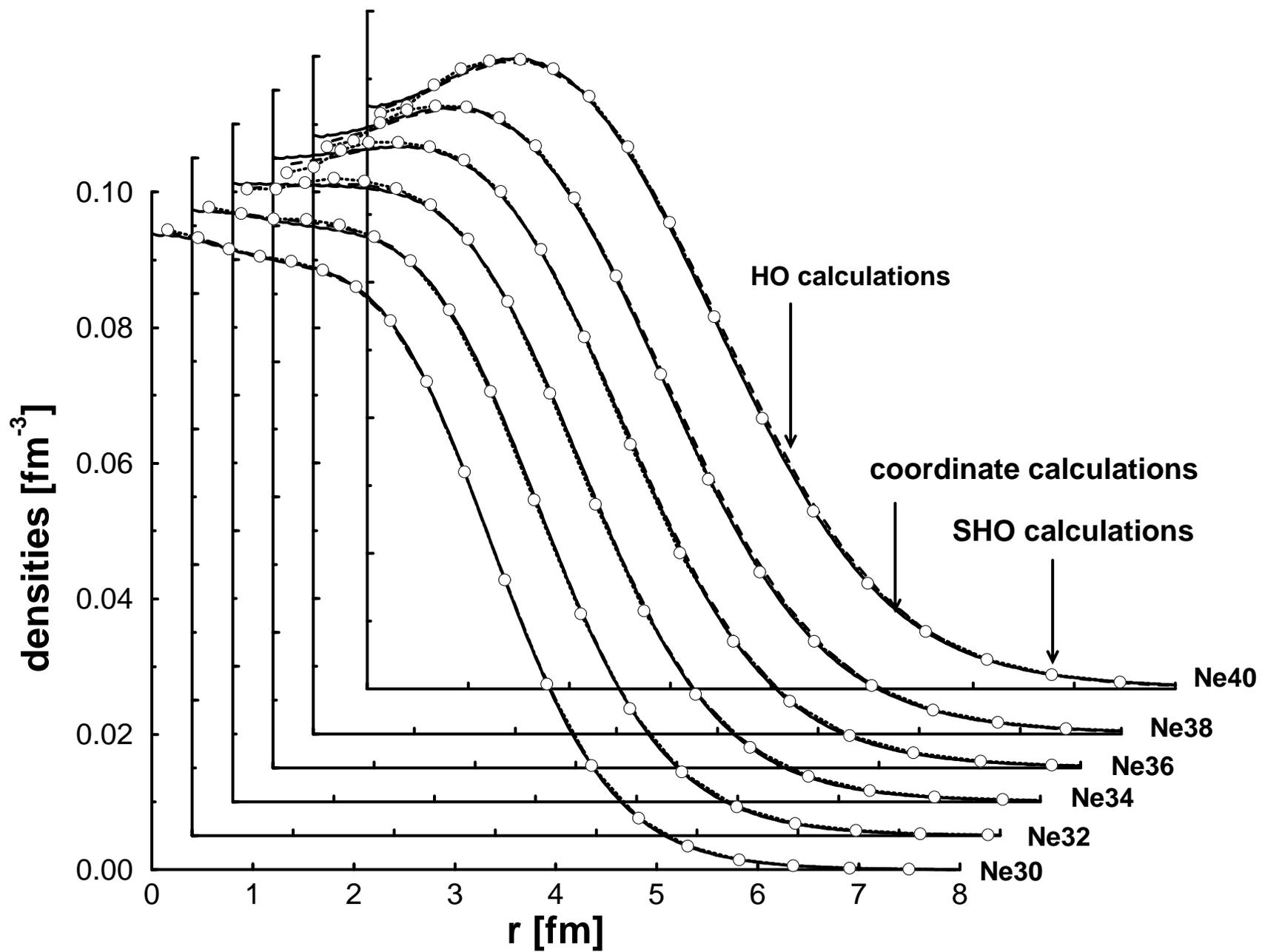
\begin{figure}
\caption{Self-consistent neutron densities for the $^{30-40}$Ne isotopes.
Distributions calculated by expansions in the HO and SHO bases are
displayed together with profiles obtained on a mesh in coordinate space.}   
\label{figC}
\end{figure}

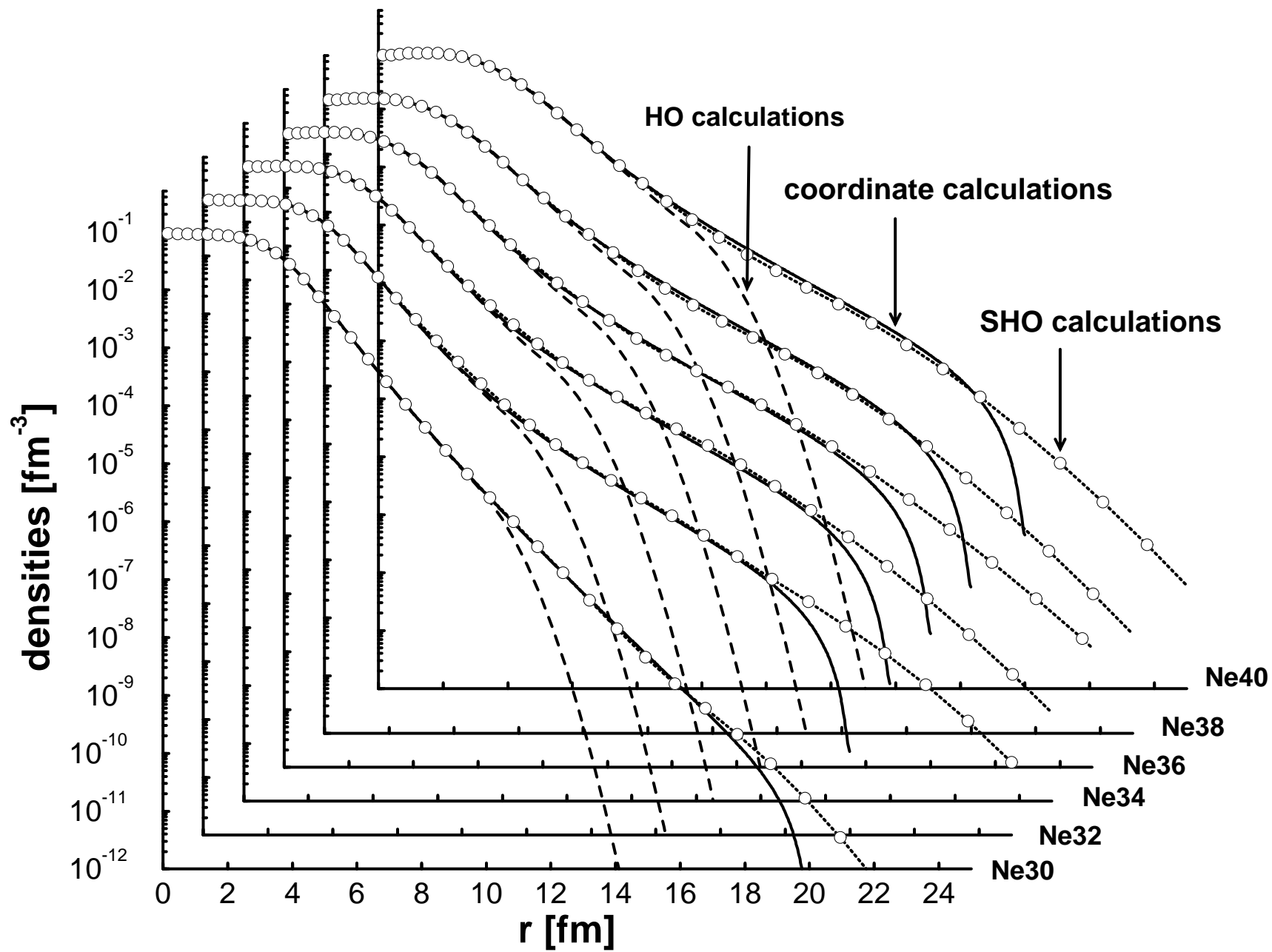
\begin{figure}
\caption{Same as in Fig. 3, but in logarithmic scale.}   
\label{figD}
\end{figure}

In Fig.~\ref{figB} we display the proton and neutron density
profiles for $^{30}$Ne and $^{40}$Ne. In our calculation
$^{40}$Ne is the last bound isotope. The proton density distributions
practically do not change with the mass number,
but the neutron densities display
an abrupt change between $^{30}$Ne and $^{32}$Ne. A long tail
reveals the formation of a multi-particle halo. Again we notice that
calculations in the HO basis do not reproduce the asymptotic 
behavior of the neutron densities. Except for the radii shown 
for comparison in Fig.~\ref{figA}, we have used a HO basis of
20 shells. While the results of HO calculations display a 
typical Gaussian dependence at large distances from the center
of the nucleus, SHO results with the same number of oscillator
shells nicely reproduce the coordinate space calculations.
More details are shown in Figs.~\ref{figC} and ~\ref{figD},
where neutron density profiles are plotted for the 
even isotopes $^{30-40}$Ne. While all three methods 
(coordinate space, HO and SHO bases) produce almost identical
density distributions for neutrons in the "inner" region $r\le 10~fm$
(Fig.~\ref{figC}), results are very different for the asymptotic behavior 
(Fig.~\ref{figD}). Not only is the SHO basis obviously
superior when compared with results of the HO calculations, but 
to some extent it could have also an advantage over calculations in 
coordinate space. Since these are performed in a box of finite
dimension ($20~fm$ in the present calculations), near the boundary
of the box a typical, very steep decrease in the densities 
is observed. Coordinate space solutions in this region of course
have nothing to do with physical densities, rather they are
an artifact of the finite dimension of the mesh. SHO solutions 
do not suffer from this drawback, and the corresponding
asymptotic densities are more natural. Of course one can increase
the coordinate mesh, but this might not always be
feasible, especially in more than one dimension.

Calculations in the configurational space generally depend on the
choice of parameters that determine the basis functions. 
In non relativistic self-consistent models the standard procedure
is to minimize the energy of the ground-state with respect to the 
basis parameters~\cite{VOT.73}. The minimization is performed in 
each step of the self-consistent iteration. This method however 
cannot be applied in relativistic models. Due to the presence 
of the Dirac sea, the energy functional does not display a 
minimum, but rather a saddle point in the ground state. The usual
approach then is to find regions of values of basis parameters 
for which all calculated quantities saturate, i.e. properties of
the ground state become independent on the specific choice of the 
basis. In particular, the spherical HO basis is 
characterized by the oscillator length parameter
$b=\sqrt{\hbar /m\omega }$. Calculations
along the $\beta $-stability line have shown that nuclear properties
saturate for $\hbar \omega \simeq 41A^{1/3}$~\cite{GRT.90}.
The situation is different for nuclei at neutron drip lines.
Properties associated with the halo states (as e.g. neutron densities
and rms radii) depend linearly on the oscillator length parameter.
The new SHO basis is determined by the two parameters: 
oscillator length $b$ and the matching radius $a$. Our calculations 
have shown that quantities which characterize the ground-state 
saturate for those values of $a$ and $b$ for which 
the neutron densites decrease exponentially 
outside the nuclear volume. 

In Ref.~\cite{PVL.97} we have shown that
the microscopic origin of the neutron halo can be found 
in a  delicate balance between the self-consistent mean-field
and pairing correlations. 
The RHB ground-state wave function can be written
either in the quasiparticle basis as a product of
independent quasi-particle states, or in the canonical basis
as a highly correlated BCS-state. The canonical states
are eigenstates of the RHB density matrix, and the eigenvalues are 
the corresponding occupation numbers. Since the density matrices
in RHB are always localized, it follows that single-particle
wave functions in the canonical basis are localized. 
The formation of the neutron halo is related to the
quasi-degeneracy of the triplet of states 1f$_{7/2}$, 2p$_{3/2}$
and  2p$_{1/2}$. The pairing interaction promotes neutrons from the
1f$_{7/2}$ orbital to the 2p levels. Due to their small centrifugal barrier,
the 2p$_{3/2}$ and 2p$_{1/2}$ orbitals form the halo. In the present work
we have also verified that calculations in the configurational
representation with the SHO basis reproduce the energies and
occupation probabilities of canonical neutron states calculated
on the mesh in coordinate space.

\section{Summary}

In the present work we have applied the relativistic Hartree-Bogoliubov
theory in the description of ground-state properties of neutron-rich
Ne isotopes. The model has been applied in the self-consistent
mean-field approximation. Pairing correlations and the coupling 
to particle continuum states are described by the pairing 
part of the finite range Gogny interaction. In many applications 
of HFB and RHB it has been emphasized that, for a correct description
of the asymptotic behavior of nucleonic densities, coordinate
space solutions for the equations that describe the
coupling between bound and continuum states are essential. 
This is especially important on the neutron-rich side of the
valley of $\beta$-stability, where extremely diffuse and spatially
extended neutron densities are found at the drip lines. For these
nuclei solutions based on the expansion of wave functions and 
mean-field potentials in harmonic oscillator bases result 
in ground-state densities with wrong asymptotic properties.
Coordinate space calculations are relatively simple in one dimension, 
i.e. for spherical nuclei. In more than one dimension, numerical
solutions on a coordinate mesh can become extremely involved, 
especially if finite range two-body interactions are 
considered. In future applications of the RHB theory to 
deformed exotic nuclei it would therefore be useful to have
alternative methods for the solution of the coupled system
of partial differential Dirac-Hartree-Bogoliubov and
Klein-Gordon equations. 

We have investigated properties of a new spherical basis
in configuration space. 
This scaled harmonic oscillator basis is derived by
a local scaling-point transformation of the spherical
harmonic oscillator radial functions. The unitary scaling
transformation produces a basis with
improved asymptotic behavior, while it preserves
useful properties of the HO wave functions.
The SHO basis has been employed in RHB calculations of
ground-state properties of Ne isotopes. Results obtained
in configurational representation have been compared 
with rms radii and densities calculated by finite element 
discretization in the coordinate space. It has been shown
that densities calculated in the SHO basis nicely reproduce
results obtained on the coordinate mesh. In particular,
using a rather small number of oscillator shells, we have
been able to reproduce the sharp increase in neutron
rms radii which had been interpreted as evidence for the
formation of a multi-particle halo. Neutron densities 
calculated in the SHO basis display the correct exponential
decrease in the asymptotic region at large distances from
the center of the potential. The present investigation has demonstrated
that SHO bases can be used in the description of the
structure of spatially extended nuclei. This might be especially
important in studies of deformed halos, where coordinate
space discretization algorithms become technically very
complicated. 
\bigskip
\begin{center}
{\bf ACKNOWLEDGEMENTS}
\end{center}

This work has been supported in part by the Bulgarian National Foundation for
Scientific Research under project $\Phi $-527, and by the
Bundesministerium f\"ur Bildung und Forschung under
project 06 TM 875.

\newpage

\centerline{\bf Figure Captions}
\bigskip

\begin{figure}
\caption{Calculated proton and neutron rms radii for the Ne isotopes.
Values obtained by finite element discretization on the coordinate
mesh are compared with calculations in the configurational
representation: expansion in the harmonic oscillator (HO) and 
scaled harmonic oscillator (SHO) bases.}
\label{figA}
\end{figure}

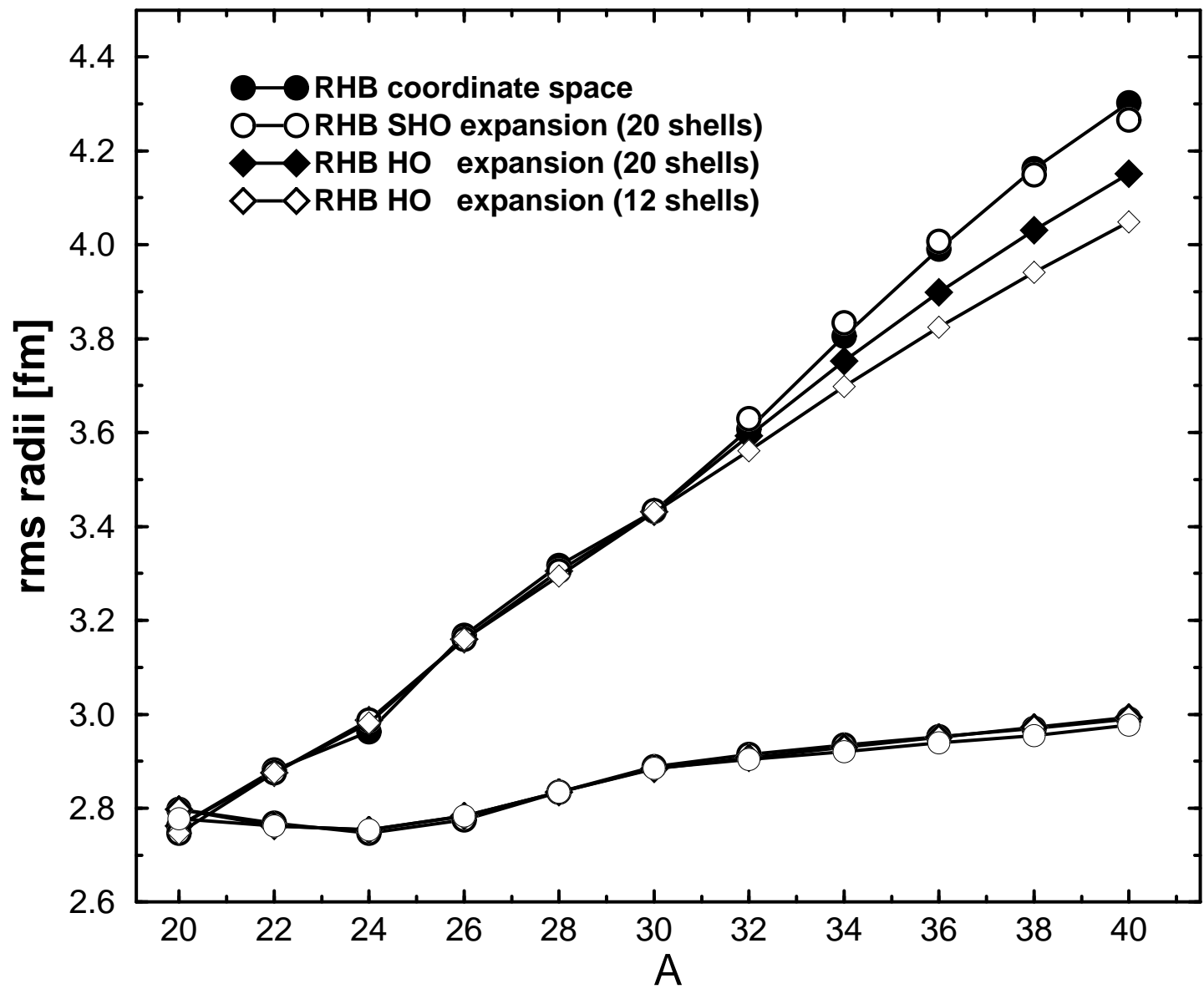
\end{document}